\documentclass[
superscriptaddress,
twocolumn,
amsmath,amssymb,
aps,
prl,
nofootinbib,
]{revtex4-2}

\makeatletter
\def\clearfmfn{\let\@FMN@list\@empty}    
\makeatother

\usepackage{comment}
\usepackage{tabularx}
\usepackage{graphicx}
\usepackage{dcolumn}
\usepackage{bm}
\usepackage{mathtools}
\usepackage{color}
\newcommand{\ket}{\rangle}
\newcommand{\bra}{\langle}

\usepackage{here}

\usepackage{hyperref}

\usepackage{xcolor}
\hypersetup{
	colorlinks,
	linkcolor={blue!70!black},
	citecolor={blue!70!black},
	urlcolor={blue!70!black}
}

\begin{document}

\title{Quantum Control of Trapped Polyatomic Molecules for eEDM Searches}

\author{Lo\"{i}c Anderegg}
\email{anderegg@g.harvard.edu}
\affiliation{Department of Physics, Harvard University, Cambridge, MA 02138, USA}
\affiliation{Harvard-MIT Center for Ultracold Atoms, Cambridge, MA 02138, USA}

\author{Nathaniel B. Vilas}
\affiliation{Department of Physics, Harvard University, Cambridge, MA 02138, USA}
\affiliation{Harvard-MIT Center for Ultracold Atoms, Cambridge, MA 02138, USA}

\author{Christian Hallas}
\affiliation{Department of Physics, Harvard University, Cambridge, MA 02138, USA}
\affiliation{Harvard-MIT Center for Ultracold Atoms, Cambridge, MA 02138, USA}

\author{Paige Robichaud}
\affiliation{Department of Physics, Harvard University, Cambridge, MA 02138, USA}
\affiliation{Harvard-MIT Center for Ultracold Atoms, Cambridge, MA 02138, USA}

\author{Arian Jadbabaie}
\affiliation{Division of Physics, Mathematics, and Astronomy, California Institute of Technology, Pasadena, CA 91125, USA}

\author{John M. Doyle}
\affiliation{Department of Physics, Harvard University, Cambridge, MA 02138, USA}
\affiliation{Harvard-MIT Center for Ultracold Atoms, Cambridge, MA 02138, USA}

\author{Nicholas R. Hutzler}
\email{hutzler@caltech.edu}
\affiliation{Division of Physics, Mathematics, and Astronomy, California Institute of Technology, Pasadena, CA 91125, USA}

\date{\today}

\begin{abstract}
Ultracold polyatomic molecules are promising candidates for experiments in quantum science, quantum sensing, ultracold chemistry, and precision measurements of physics beyond the Standard Model. A key, yet unrealized, requirement of these experiments is the ability to achieve full quantum control over the complex internal structure of the molecules. Here, we establish coherent control of individual quantum states in a polyatomic molecule, calcium monohydroxide (CaOH), and use these techniques to demonstrate a method for searching for the electron electric dipole moment (eEDM). Optically trapped, ultracold CaOH molecules are prepared in a single quantum state, polarized in an electric field, and coherently transferred into an eEDM sensitive state where an electron spin precession measurement is performed. To extend the coherence time of the measurement, we utilize eEDM sensitive states with tunable, near-zero magnetic field sensitivity. The spin precession coherence time is limited by AC Stark shifts and uncontrolled magnetic fields. These results establish a path for eEDM searches with trapped polyatomic molecules, towards orders-of-magnitude improved experimental sensitivity to time-reversal-violating physics.
\end{abstract}

\maketitle

The rich structure of polyatomic molecules makes them an appealing platform for experiments in quantum science~\cite{wall2013simulating,wall2015realizing,wei2011entanglement,yu2019scalable}, ultracold chemistry~\cite{augustovicova2019collisions}, and precision measurements~\cite{kozyryev2017precision,kozyryev2021enhanced,hutzler2020polyatomic,norrgard2019nuclear,hao2020nuclear}. Key to this structure is the presence of near-degenerate states of opposite parity, which allow the molecules to be easily polarized in the laboratory frame with the application of a small electric field. Such states are a novel resource, generic among polyatomic molecules while rare in diatomics, that may be useful for applications such as analog simulation of quantum magnetism models~\cite{wall2013simulating,wall2015realizing} or for realizing switchable interactions and long-lived qubit states for quantum computing~\cite{yu2019scalable}. Additionally, the parity-doublet states in trapped polyatomic molecules are expected to be an invaluable tool for systematic error rejection in precision measurements of physics beyond the Standard Model (BSM)~\cite{kozyryev2017precision}. To date, several species of polyatomic molecules have been laser cooled and/or trapped at ultracold temperatures~\cite{kozyryev2017sisyphus,augenbraun2019laser,mitra2020direct,vilas2022magneto,hallas2022optical,zeppenfeld2012sisyphus,prehn2016optoelectrical}. 

One powerful avenue for tabletop BSM searches is probing for the electric dipole moment of the electron (eEDM)~\cite{demille2017Probing,chupp2019edm,safronova2018search,Cesarotti2019,Pospelov2005}, $d_e$, which violates time-reversal (T) symmetry and is predicted by many BSM theories to be orders of magnitude larger than the Standard Model prediction~\cite{safronova2018search,chupp2019edm}. Current state-of-the-art eEDM experiments are broadly sensitive to T-violating physics at energies much greater than 1~TeV~\cite{Hudson2011,baron2014order,acme2018improved,cairncross2017precision,roussy2022new,alarcon2022edm}. All such experiments use Ramsey spectroscopy to measure an energy shift due to the interaction of the electron with the large electric field present inside a polarized molecule~\cite{baron2014order,acme2018improved,cairncross2017precision,zhou2020second,roussy2022new}. Molecular beam experiments have achieved high statistical sensitivity by measuring a large number of molecules over a $\approx$~$1~$ms coherence time~\cite{baron2014order,acme2018improved}, while molecular ion-based experiments have used long Ramsey interrogation times ($\approx$~$1~$s) though with lower numbers~\cite{cairncross2017precision,zhou2020second,roussy2022new}. Measurements with trapped neutral polyatomic molecules can potentially combine the best features of each approach to achieve orders-of-magnitude improved statistical sensitivity~\cite{kozyryev2017precision}.

In this Report, we demonstrate full quantum control over the internal states of a trapped polyatomic molecule in a vibrational bending mode with high polarizability in small electric fields. The method starts with preparing ultracold, optically trapped molecules in a single hyperfine level, after which a static electric field is applied to polarize the molecules. The strength of the polarizing electric field is tuned to obtain near-zero \textit{g}-factor spin states, which have strongly suppressed sensitivity to magnetic field noise while retaining eEDM sensitivity. Microwave pulses are applied to create a coherent superposition of these zero \textit{g}-factor spin states that precesses under the influence of an external magnetic field. The precession phase is then read out by a combination of microwave pulses and optical cycling. 

\begin{figure*}[t]
    \resizebox{0.94\textwidth}{!}
    {\includegraphics{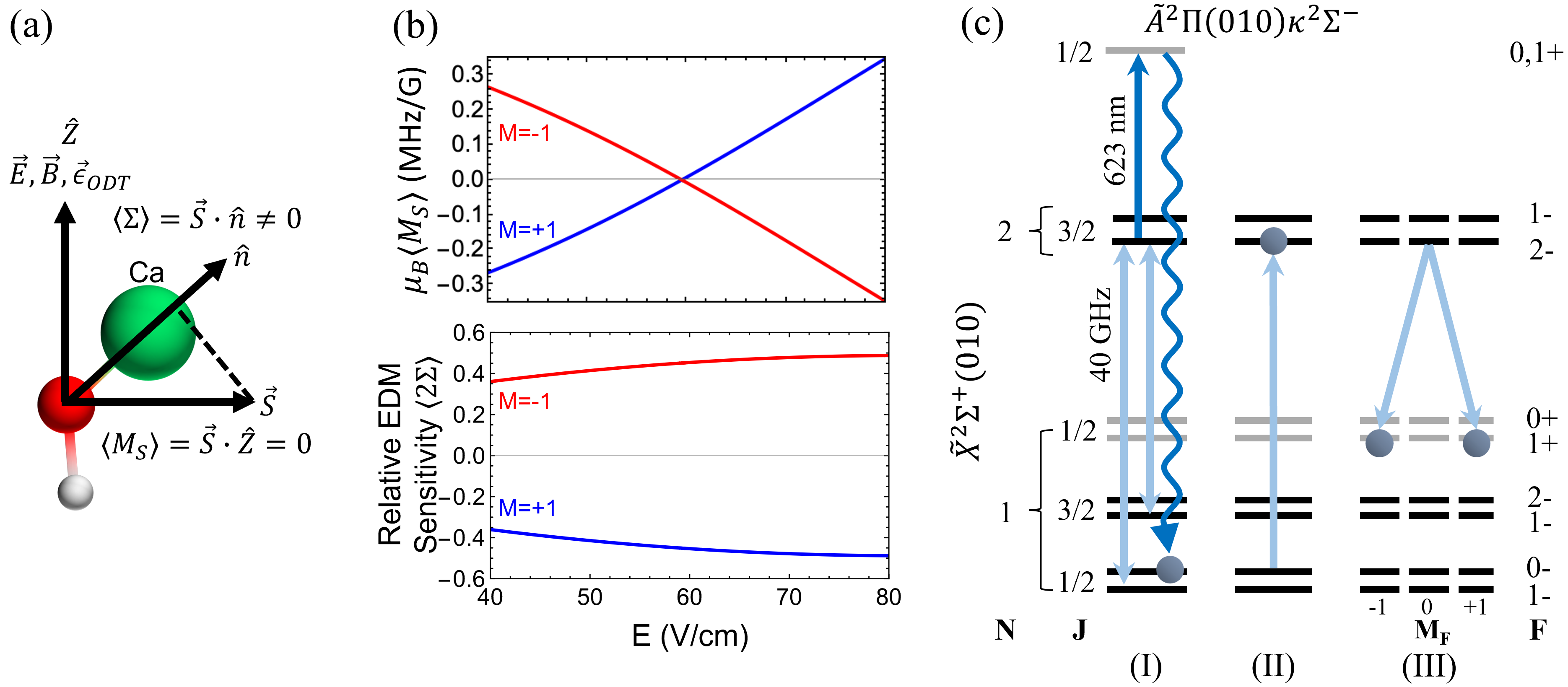}}
    \caption{\label{fig1}  (a) A geometric picture of the bending molecule at the zero \textit{g}-factor crossing, showing the electron spin ($\vec{S}$) has a finite projection on the molecule axis ($\hat{n}$), giving eEDM sensitivity. However, the electron spin ($\vec{S}$) is orthogonal to the magnetic field ($\vec{B}$), resulting in suppressed magnetic field sensitivity. (b) The magnetic sensitivity (upper plot) and eEDM sensitivity (lower plot) for a pair of zero \textit{g}-factor states ($N=1, J=1/2^+, F=1, M_F=\pm1$) are shown as a function of the applied electric field. (c) Experimental sequence to prepare the eEDM sensitive state. First, the molecules are pumped into a single quantum state ($N=1, J=1/2^-, F=0$) with a combination of microwave drives and optical pumping (I). Next, a microwave $\pi$-pulse drives the molecules into the $N=2,J=3/2^-, F=2, M_F=0$ state (II). Lastly, the eEDM measurement state is prepared as a coherent superposition of the $N=1, J=1/2^-, F=1$ $M_F=\pm1$ states with a microwave $\pi$-pulse (III). The states which are optically detectable with the detection light are shown in black, while those not addressed by the detection light are in grey.}
\end{figure*}

We observe spin precession over a range of electric and magnetic fields and characterize the current limitations to the coherence time of the measurement. With readily attainable experimental parameters, coherence times on the order of the state lifetime ($>$100~ms) could be realistically achieved. We therefore realize the key components of an eEDM measurement in this system. Although the light mass of CaOH precludes a competitive eEDM measurement~\cite{Gaul2020Triatomic}, the protocol demonstrated here is directly transferable to heavier laser-cooled alkaline earth monohydroxides with identical internal level structures, such as SrOH, YbOH, and RaOH, which have significantly enhanced sensitivity to the eEDM~\cite{kozyryev2017sisyphus,kozyryev2017precision,Isaev2017RaOH,augenbraun2019laser,Gaul2020Triatomic}.

\newcommand{\Eeff}{\mathcal{E}_\text{eff}}

In eEDM measurements with polarized molecules, the electron spin $\vec{S}$ precesses under the influence of an external magnetic field $B_Z$ and the internal electric field of the molecule, $\Eeff$, which can be large due to relativistic effects. Time evolution is described by the Hamiltonian
\begin{eqnarray}
    H & = & g_{S} \mu_B B_{Z} 
    \vec{S} \cdot\hat{Z} - d_e \Eeff \vec{S} \cdot \hat{n} \nonumber \\
    & = & g_{S} \mu_B B_{Z} 
    M_S - d_e \Eeff \Sigma.
    \label{eqn:ham}
\end{eqnarray}\\
Here, $g_S\approx 2$ is the electron spin \textit{g}-factor, $\mu_B$ is the Bohr magneton, $B_Z$ points along the lab $\hat{Z}$ axis, and the internal field $\Eeff$ points along the molecule's internuclear axis $\hat{n}$. We define the quantities $M_S=\vec{S}\cdot \hat{Z}$ and $\Sigma = \vec{S}\cdot \hat{n}$ to describe the electron's magnetic sensitivity and EDM sensitivity, respectively. The effect of the eEDM can be isolated by switching the orientation of the applied magnetic field or, alternatively, by switching internal states to change the sign of $M_S$ or $\Sigma$. Performing both switches is a powerful technique for suppressing systematic errors~\cite{acme2018improved,cairncross2017precision}.

\begin{figure*}[t]
    \resizebox{.8\textwidth}{!}{\includegraphics{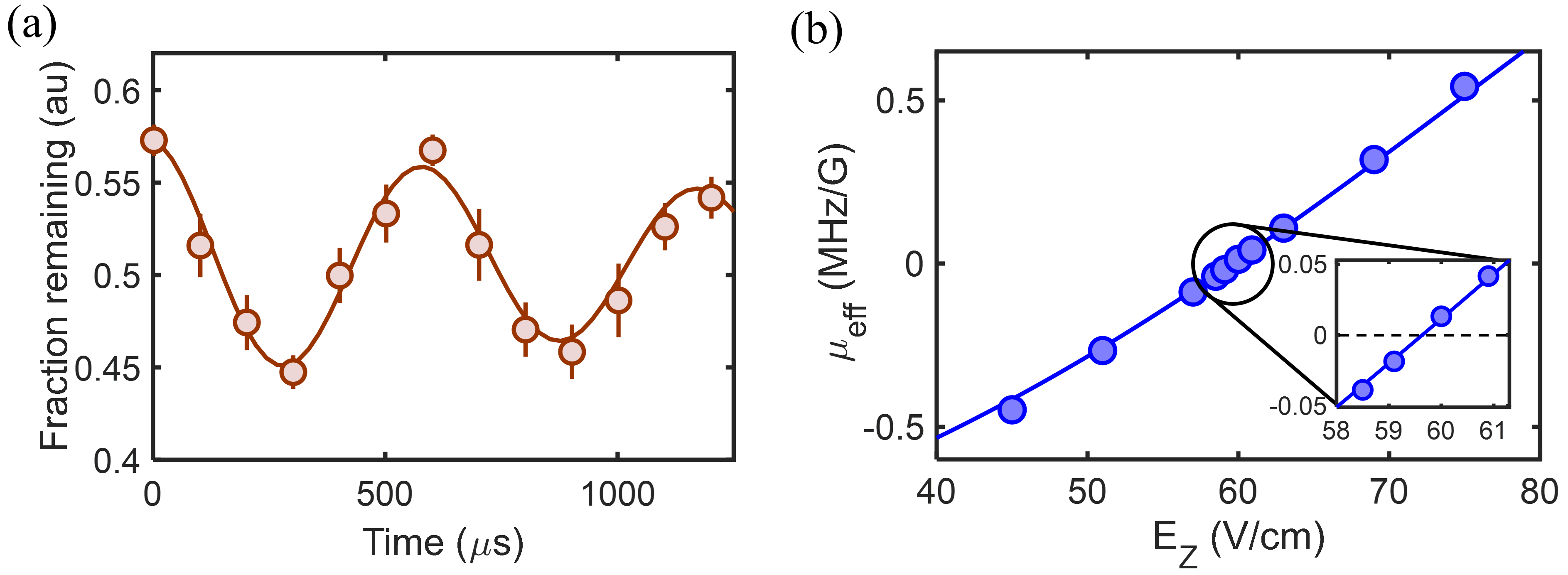}}
    \caption{(a) Spin precession of the eEDM sensitive state in the presence of a bias magnetic field.  (b) Magnetic field sensitivity of the eEDM state in CaOH as a function of electric field. The field sensitivity is determined by measuring the spin precession frequency at different electric fields with an applied magnetic field of $B_Z = 110~\text{mG}$. Error bars are smaller than the markers. The solid curve is the calculated magnetic field sensitivity in the presence of trap shifts using known molecular parameters, as described in the Supplemental Material.}
    \label{fig2}
\end{figure*}

Current EDM bounds rely on specific states in diatomic molecules that have an unusually small \textit{g}-factor, reducing sensitivity to stray magnetic fields~\cite{baron2014order,cairncross2017precision}. However, CaOH, like other laser-coolable molecules with structure amenable to eEDM searches~\cite{Isaev2017RaOH,kozyryev2017precision,isaev2016polyatomic,kozyryev2016proposal}, has a single valence electron, which results in large magnetic \textit{g}-factors. In this work, we engineer reduced magnetic sensitivity by using an applied electric field $E_Z$ to tune $M_S$ to a zero-crossing, while maintaining significant eEDM sensitivity $\Sigma$. This technique is generic to polyatomic molecules with parity-doublets. Details of a specific $M=\pm 1$ pair of zero \textit{g}-factor states are shown in Figure \ref{fig1} (a)-(b), with further information in the Supplemental Material. Sensitivity to transverse magnetic fields is also suppressed in these zero \textit{g}-factor states (see Supplemental Material). 

The experiment begins with laser-cooled CaOH molecules loaded from a magneto-optical trap \cite{vilas2022magneto} into an optical dipole trap (ODT) formed by a 1064 nm laser beam with a 25 $\mu$m waist size, as described in previous work \cite{hallas2022optical}. The ODT is linearly polarized and its polarization vector $\vec{\epsilon}_\text{ODT}$ defines the $\hat{Z}$ axis, along which we also apply magnetic and electric fields, $\vec{B} = B_Z \hat{Z}$ and $\vec{E} = E_Z \hat{Z}$, respectively, as depicted in Figure 1(a). We first non-destructively image the molecules in the ODT for 10 ms as normalization against variation in the number of trapped molecules. The molecules are then optically pumped into the $N=1^-$ levels of the $\widetilde{X}{}^2\Sigma^+(010)$ vibrational bending mode \cite{hallas2022optical} (Figure 1(c)), and the trap depth is adiabatically lowered by $3.5\times$ to reduce the effect of AC Stark shifts from the trap light and to lower the temperature of the molecules to 34 $\mu$K. Any molecules that were not pumped into $N=1^-$ levels of the bending mode are heated out of the trap with a pulse of resonant laser light.

Following transfer to the $\widetilde{X}{}^2\Sigma^+(010)(N=1^-)$ state, the molecular population is initially spread across twelve hyperfine Zeeman sublevels in the spin-rotation components $J=1/2$ and $J=3/2$. To prepare the molecules in a single hyperfine state, we use a combination of optical pumping and microwave pulses, as shown in Figure~\ref{fig1}(c). We first apply microwaves from the $(N=1, J=3/2^-)$ state up to the $(N=2, J=3/2^-)$ state. As this transition is parity-forbidden, we apply a small electric field $E_Z = 7.5$~V/cm to slightly mix the parity of the $N=1$ levels and provide transition strength. From the $N=2$ state, we drive an optical transition to the excited $\widetilde{A}^2\Pi (010) \kappa {}^2 \Sigma^{(-)}, J=1/2^+$ state. This state predominately decays to both $F=0$ (the target state) and $F=1$ states in the $N=1, J=1/2^-$ manifold. After 3 ms of optical pumping, the microwaves are switched to drive the accumulated $N=1, J=1/2^-, F= 1$ population to the same $N=2, J=3/2^-$ state in $\widetilde{X}(010)$, where they are excited by the optical light and pumped into the target $F=0$ state. Once this optical pumping sequence is complete, we adiabatically ramp the electric field to $E_Z=$150~V/cm to significantly mix parity, then drive population up to the $N=2, J=3/2^-, F=2, M=0$ state with a microwave $\pi$-pulse (Figure \ref{fig1}(c)(II)). We clean out any remaining population in the $N=1$ state with a depletion laser that resonantly drives population to undetected rotational levels. 

\begin{figure*}[t]
    \resizebox{\textwidth}{!}{\includegraphics{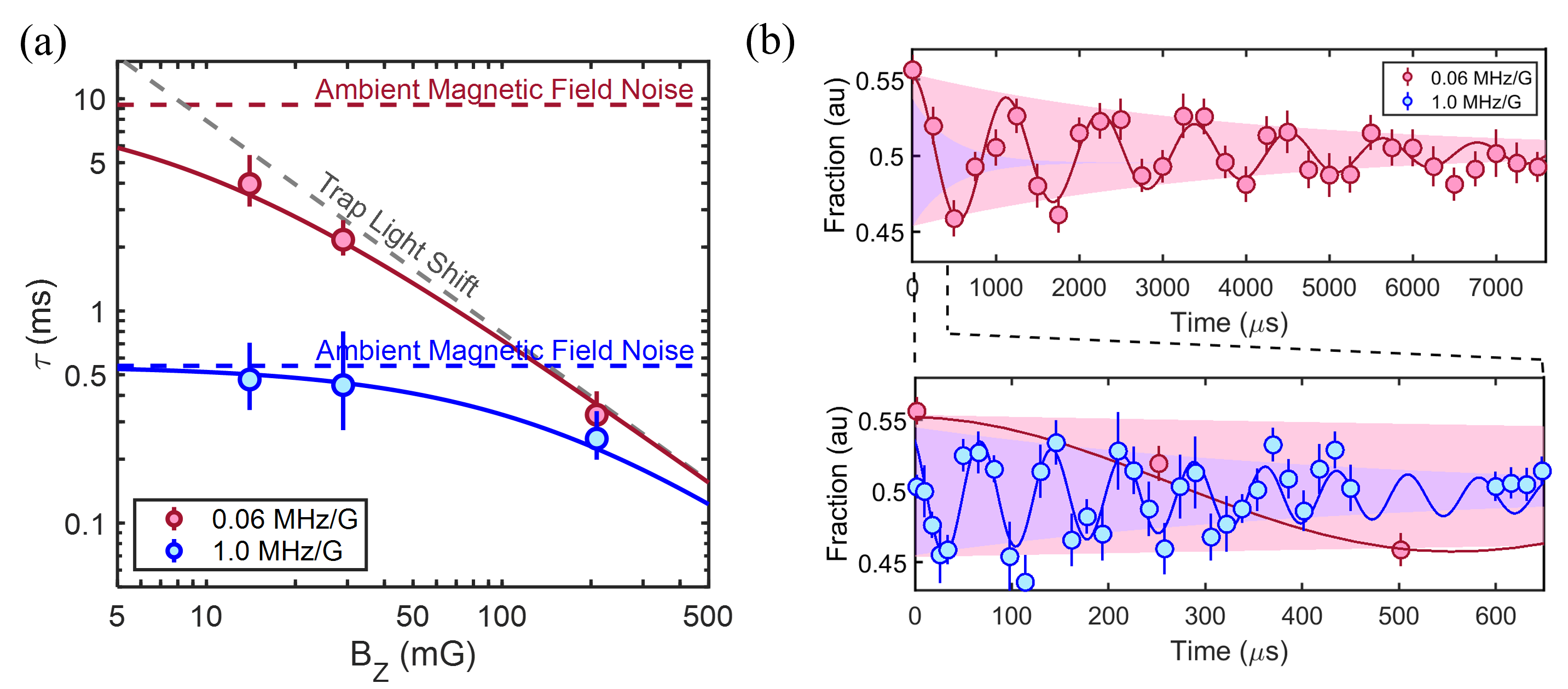}}
    \caption{Coherence time of the spin precession signal. (a) Measured coherence times $\tau$ versus $B_\text{Z}$ at different electric fields (red and blue markers, corresponding to different magnetic field sensitivity). The coherence time scales as $1/B_\text{Z}$ due to AC Stark shift broadening, then plateaus at a limit set by the magnetic field instability $\delta B$. This limit increases as the $g$-factor approaches zero. Solid and dashed curves are fit to the data. The ambient magnetic field noise determined from the fit is $\delta B = 4_{-1}^{+2}$ mG, while the fitted decoherence time due to light shifts is $\tau = (1/B_Z)\times80_{-10}^{+20}$ ms$\times$mG. (b) The spin precession coherence time at $B_Z = 15$~mG is extended by 16$\times$ by approching the zero \textit{g}-factor point.}
    \label{fig3}
\end{figure*}

\begin{figure}[t]
    \resizebox{.48\textwidth}{!}{\includegraphics{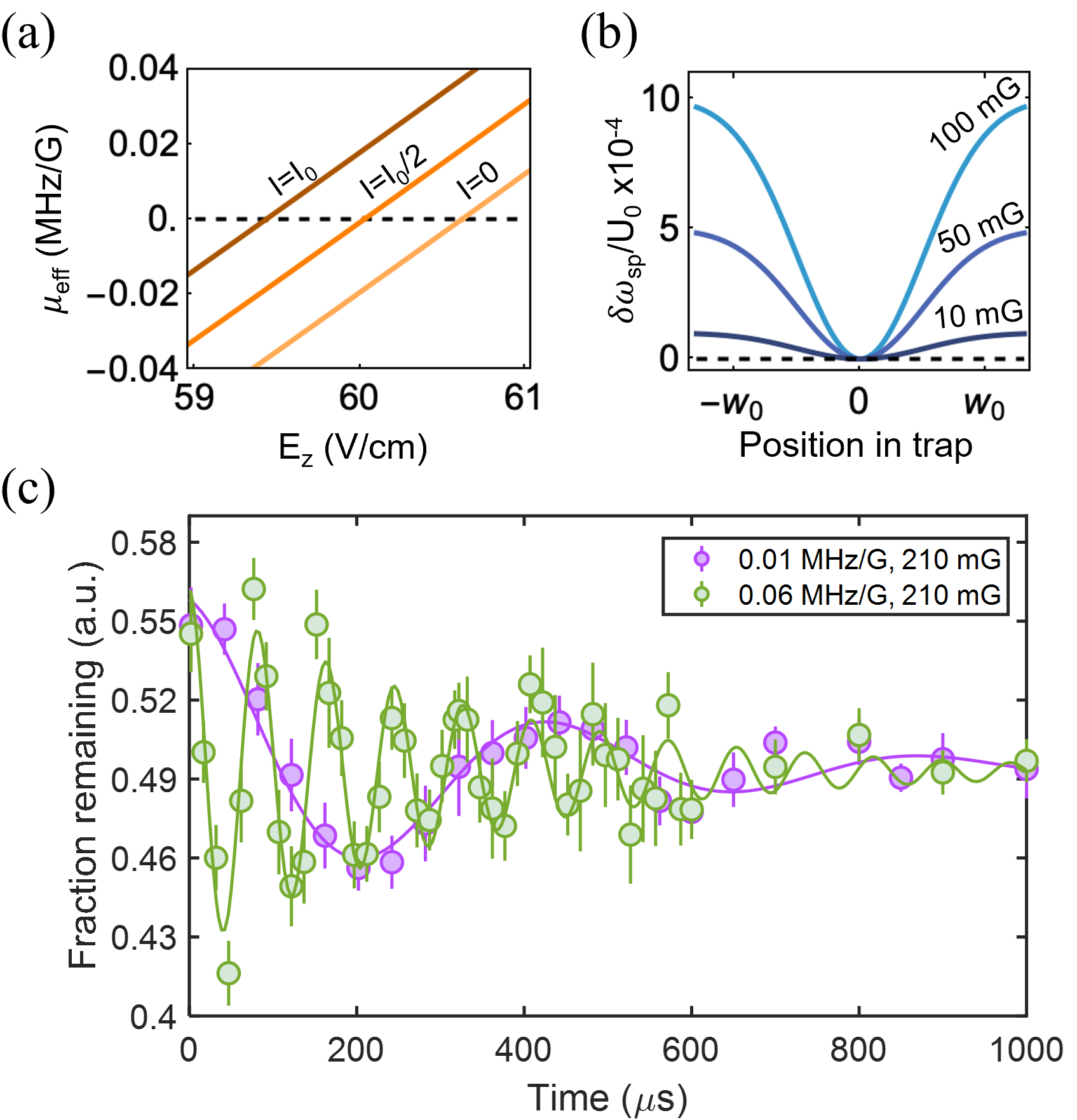}}
    \caption{Effect of trap light on coherence time. (a) The trap light shifts the location of the zero crossing in $\mu_\text{eff}$. As a result, molecules at a finite temperature explore different magnetic field sensitivities $\mu_\text{eff}$. (b) Dependence of the spin precession frequency (scaled by the trap depth $U_0$) on position within the trap. At lower magnetic fields, the relative change in spin precession frequency is reduced.  (c) Two spin precession curves taken at the same magnetic field ($B_Z = 210$~mG) but at different electric fields, showing that the AC Stark shift limitation is independent of the effective \textit{g}-factor, since AC Stark shifts dominate the coherence time for large bias fields.}
     \label{fig4} 
\end{figure}

To perform spin precession in the eEDM sensitive state, we first adiabatically ramp the electric field to a value $E_Z$, then turn on a small bias magnetic field $B_Z$. We measure the electron spin precession frequency using a procedure analogous to Ramsey spectroscopy \cite{acme2018improved, baron2014order}. The molecules are prepared by driving a $\pi$-pulse (2.5 $\mu$s), with microwaves linearly polarized along the lab $\hat{X}$ axis, into the ``bright'' superposition state $|B\rangle = (|M=1\rangle + |M=-1\rangle)/\sqrt{2}$ within the $N=1, J=1/2^+, F=1, M=\pm1$ eEDM sensitive manifold (Figure \ref{fig1}(c)). The state begins to oscillate between the bright state and the ``dark'' state $|D\rangle = (|M=1\rangle - |M=-1\rangle)/\sqrt{2}$ at a rate $\omega_\text{SP} = \mu_\text{eff}B_Z$, where the effective magnetic moment $\mu_\text{eff}=\mu_B g_\text{eff}=g_S \mu_B (\langle M_S \rangle_{M=1}-\langle M_S \rangle_{M=-1})$ is tuned via the applied electric field $E_\text{Z}$ (Figure \ref{fig1}(b)). The contribution from the $d_e \mathcal{E}_\text{eff}$ term in eqn. \ref{eqn:ham} is negligible in CaOH, but could be measured in heavier molecules with much larger $\Eeff$. After a given time, a second $\pi$-pulse is applied to stop spin precession and transfer the bright state to the optically detectable $N=2, J=3/2^-$ level.
Once the electric field is ramped down, the population remaining in the eEDM manifold, which has the opposite parity, is not optically detectable. We then image the ODT again and take the ratio of the first and second images (Figure \ref{fig2}(a)). At long spin precession times ($>10$ ms), losses from background gas collisions ($\sim$1~sec), blackbody excitation ($\sim$1~sec), and the spontaneous lifetime of the bending mode ($\sim$0.7~sec) lead to an overall loss of signal, as characterized in Ref.~\cite{hallas2022optical}. This effect is mitigated with a fixed duration between the first and second images, making the loss independent of the precession time.

To map out the location of the zero \textit{g}-factor crossing, we perform spin precession measurements at a fixed magnetic field $B_Z = 110$ mG for different electric fields (Figure \ref{fig2}(b)). The spin precession frequency corresponds to an effective \textit{g}-factor at that electric field. We find that the zero \textit{g}-factor crossing within the $N=1, J=1/2^+, F=1, M=\pm1$ eEDM manifold occurs at an electric field of 59.6 V/cm, in agreement with theory calculations described in the Supplemental Material. We note that there is another zero \textit{g}-factor crossing for the $N=1, J=3/2^+, F=1$ manifold at $\approx$ 64 V/cm, which has a smaller eEDM sensitivity but the opposite slope of $g_\text{eff}$ vs. $E_Z$, thereby providing a powerful resource to reject systematic errors related to imperfect field reversals (see Supplemental Material). We emphasize that while the location of these crossings is dependent on the structure of a specific molecule, their existence is generic in polyatomic molecules, which naturally have parity-doublet structure~\cite{kozyryev2017precision}.

A critical component of the spin precession measurement is the coherence time, which sets the sensitivity of an eEDM search. Figure 3(a) shows the measured coherence time of our system at different applied fields $B_Z$ and $E_Z$. We characterize two dominant limitations that wash out oscillations at long times. Variations in the spin precession frequency can be linearly expanded as $\delta \omega_\text{SP} = \mu_\text{eff} (\delta B_Z) + (\delta \mu_\text{eff}) B_Z$. The first term describes magnetic field noise and drift of the applied bias field, given by $\delta B_Z$. The second term describes noise and drifts in the $g$-factor, $\delta g_\text{eff}$, which can arise from instability in the applied electric field, $E_Z$, or from AC Stark shifts (described below). Drifts in the bias electric field $E_Z$ are found to be negligible in our apparatus.

Decoherence due to magnetic field noise, $\delta B_Z$, is independent of the applied magnetic field but is proportional to $\mu_\text{eff}$, and can be mitigated by operating near the zero \textit{g}-factor crossing. As shown in Fig. 3(b), at an electric field of 90 V/cm, corresponding to a large magnetic moment of $\mu_\text{eff}= 1.0$ MHz/G, we realize a magnetic field noise-limited coherence time of 0.5 ms at $B_Z\approx 15$ mG. At an electric field of 61.5 V/cm, corresponding to $\mu_\text{eff} = 0.06$ MHz/G, much closer to the zero \textit{g}-factor location, we find a coherence time of 4 ms at the same $B_Z$.

At higher magnetic fields, the primary limitation to the coherence time is due to AC stark shifts from the optical trapping light (Fig. 4). The intense $Z$-polarized ODT light leads to a shift in the electric field at which the zero \textit{g}-factor crossing occurs. Due to the finite temperature of the molecules within the trap, they will explore different intensities of trap light and hence have different values of $g_\text{eff}$. The spread $\delta g_\text{eff}$ causes variation of $\omega_\text{SP}$, which leads to decoherence. In contrast to the magnetic field noise term, this effect is independent of the electric field $E_Z$ but decreases monotonically with $B_Z$, which scales the frequency sensitivity to \textit{g}-factor variations, $\delta \omega_\text{SP} = B_Z \delta \mu_\text{eff}$. The insensitivity of \textit{g}-factor broadening to the exact value of $g_\text{eff}$ is demonstrated in Fig. 4(c). Decoherence due to AC Stark shifts can be reduced by cooling the molecules to lower temperatures or by decreasing $B_Z$. The bias magnetic field can be reduced arbitrarily far until either transverse magnetic fields or magnetic field noise become dominant. From the decoherence rates measured in this work, it is expected that AC Stark shift-limited coherence times $\sim$1 s could be achieved at bias fields of $B_Z\sim100$ $\mu$G.

From the above discussion, it is expected that the longest achievable coherence times will occur for very small \textit{g}-factors, $g_\text{eff} \approx 0$, and very small bias fields, $B_Z \approx 0$. Minimizing $B_Z$ requires reducing the effects of both magnetic field noise and transverse magnetic fields to well below the level of the bias field energy shifts. We cancel the transverse magnetic fields to below 1 mG by maximizing the spin precession period under the influence of transverse $B$ fields only, and actively monitor and feedback on the magnetic field along each axis to minimize noise and drifts in $B_Z$. Note that the stainless steel vacuum chamber has no magnetic shielding, leading to high levels of magnetic field noise which would not be present in an apparatus designed for an eEDM search. Even under these conditions, we achieve a coherence time of 30 ms at an electric field of 60.3 V/cm (corresponding to $\mu_\text{eff}=0.02$ MHz/G) and a bias field of $B_Z\approx2$ mG, (see Supplemental Material). However, at such a low bias field, the molecules are sensitive to 60 Hz magnetic field noise present in the unshielded apparatus, which is on the same order as the bias field. Since the experiment is phase stable with respect to the AC line frequency, this 60 Hz magnetic field fluctuation causes a time-dependent spin precession frequency. Nevertheless, our prototype experiment confirms that long coherence times are possible. Any future eEDM experiment would have magnetic shielding that would greatly suppress nefarious magnetic fields from the environment. Such shielding could readily enable coherence times exceeding that of the $\sim0.5$~s lifetime of the bending modes of similar linear polyatomic molecules with larger eEDM sensitivity~\cite{hallas2022optical}.

In summary, we have realized coherent control of optically trapped polyatomic molecules and demonstrated a realistic experimental roadmap for future eEDM measurements. By leveraging the unique features of the quantum levels in polyatomic molecules, we achieve a coherence time of 30 ms for paramagnetic molecules in a stainless steel chamber with no magnetic shielding. With common shielding techniques employed in past EDM experiments, there is a clear path to reducing stray fields and extending coherence times to $>100$ ms. At such a level, the dominant limitation becomes the finite lifetime of the bending mode \cite{hallas2022optical}. Even longer coherence times are possible with the right choice of parity doublet states, as found in symmetric or asymmetric top molecules \cite{augenbraun2020molecular, mitra2020direct, kozyryev2017precision, augenbraun2021observation}. 

Following our established roadmap with heavier trapped polyatomic molecules has the potential to provide orders-of-magnitude improvements to current bounds on T-violating physics. Using a recent study of the $\widetilde{X}(010)$ state in YbOH~\cite{jadbabaie2022characterizing}, we have identified similar $N=1$ zero \textit{g}-factor states for eEDM measurements with significantly improved sensitivity.  In addition to the \textit{g}-factor tuning demonstrated in this work, polyatomic molecules provide the ability to reverse the sign of $\Sigma$ without reversing $M_S$ - a crucial feature of recent experiments that have greatly improved the limit on the eEDM~\cite{acme2018improved, roussy2022new}. For example, in the $N=1$ manifold of CaOH, there is another zero \textit{g}-factor crossing at a nearby electric field value, with 69\% smaller values of $\Sigma$ and opposite sign. Since the ratio of eEDM sensitivity to \textit{g}-factor vs. $E_Z$ slope differs between these two crossings, measurements at both points could be used to suppress systematics due to non-reversing fields coupling to the electric field dependence of the \textit{g}-factor~\cite{acme2018improved}. 

This work provides a first experimental demonstration of the advantages of the rich level structure of polyatomic molecules for precision measurements. While we have focused here on spin precession with $T$-reversed states ($M=\pm 1$), many levels of interest can be favorably engineered for precision measurement experiments. In a recent proposal~\cite{norrgard2019nuclear}, parity-doublets, magnetically tuned to degeneracy in optically trapped polyatomic molecules, were shown to be advantageous for searches for parity violating physics. In another recent work~\cite{kozyryev2021enhanced}, a microwave clock between rovibrational states in SrOH was proposed as a sensitive probe of ultra-light dark matter, utilizing transitions tuned to electric and/or magnetic insensitivity. In these proposals, and now experimentally demonstrated in our work, coherent control and state engineering in polyatomic molecules can mitigate systematic errors and enable robust searches for new physics.

This work was supported by the AFOSR and the NSF. LA acknowledges support from the HQI, NBV from the DoD NDSEG fellowship program, and PR from the NSF GRFP. NRH and AJ acknowledge support from NSF CAREER (PHY-1847550), The Gordon and Betty Moore Foundation (GBMF7947), and the Alfred P. Sloan Foundation (G-2019-12502). AJ acknowledges helpful discussions with Chi Zhang and Phelan Yu.

\clearpage
\newpage

\onecolumngrid
\begin{center}
	\textbf{\large Supplemental Material for ``Quantum Control of Trapped Polyatomic Molecules for eEDM Searches''}
\end{center}

\setcounter{figure}{0}

\makeatletter 
\renewcommand{\thefigure}{S\@arabic\c@figure}
\makeatother

\section{Zero \lowercase{\textit{g}}-factor States}

\subsection{Origin}

In ${}^2\Sigma$ electronic states of linear polyatomic molecules, the spin-rotation interaction, $\gamma \vec{N}\cdot \vec{S}$, couples the molecular rotation $N$ and the electron spin $S$ to form the total angular momentum $J$. These states are well described in the Hund's case (b) coupled basis. An applied electric field $E_Z$ will interact with the molecular-frame electric dipole moment $\mu_E$, connecting states with opposite parity, $\Delta M_F=0$, and $\Delta J \leq 1$. When $\mu_E E_Z \gg \gamma$, $N$ and $S$ are uncoupled and well described by their lab frame projections $M_N$ and $M_S$. However, in the intermediate field regime with $\mu_E E_Z \sim \gamma$, the molecular eigenstates are mixed in both the  Hund's case (b) coupled basis and the decoupled basis. $M_F$ remains a good quantum number in the absence of transverse fields. In this regime, $M_F\neq0$ states with $\langle M_S \rangle=0$ can arise at specific field values. These states have no first order electron spin magnetic sensitivity, and, unlike $M_F=0$ clock states, have large eEDM sensitivity near $B_Z=0$. We refer to these states as \textit{zero \textit{g}-factor} states~\cite{kozyryev2017precision}. 

Zero \textit{g}-factor states arise from avoided level crossings as free field states are mixed by the electric field. One of the crossing states has $\bra M_S \ket <0$, the other state has $\bra M_S \ket >0$, and both have mixed $M_N$. The spin-rotation interaction couples the states and lifts the crossing degeneracy, resulting in eigenstates that are superpositions of electron spin up and down with $\langle M_S \rangle =0$, while retaining non-zero molecular orientation with $\langle \hat{n} \rangle = \langle M_N \ell \rangle \neq0$. The lab frame projection of $\hat{n}$ ensures that the eEDM interaction in the molecule frame does not rotationally average away.

Zero \textit{g}-factor states are generically present in the Stark tuning of polyatomic molecules. The reduction of symmetry in a polyatomic molecule allows for rotation about the internuclear axis, resulting in closely spaced doublets of opposite parity. When these doublets are mixed by an applied electric field, they split into $2N+1$ groups of levels representing the values of the molecular orientation $\langle M_N\ell\rangle$. For each $N$ manifold with parity doubling, avoided level crossings generically occur between an $M_N\ell = \pm1$ Stark manifold and an $M_N\ell=0$ Stark manifold. 

In diatomic molecules without parity-doubling, the existence of zero \textit{g}-factor states requires an inverted spin rotation structure ($\gamma<0$), such that the two $J$ states are tuned closer to each other by an electric field. For example, the YbF molecule ($\gamma = -13.4$ MHz~\cite{sauer1996laser,dickinson2001fourier}) has zero \textit{g}-factor states at $E\approx 866 $ V/cm in the $N=1$ manifold, while CaF does not. However, since $|\gamma|/B\ll 1$ for most $^2\Sigma$ diatomic molecules, the electric fields that mix spin-rotation states are much less than those that polarize the molecule. Therefore, zero \textit{g}-factor states occur when the molecule has negligible lab-frame polarization, limiting eEDM sensitivity. For example, the aforementioned states in YbF have $|\langle \Sigma \rangle | \approx 0.006$, which is $\sim$3\% the value of $\Sigma$ in the zero $g$-factor states used in this work. 

\begin{figure}[b]
    \resizebox{\textwidth}{!}{\includegraphics{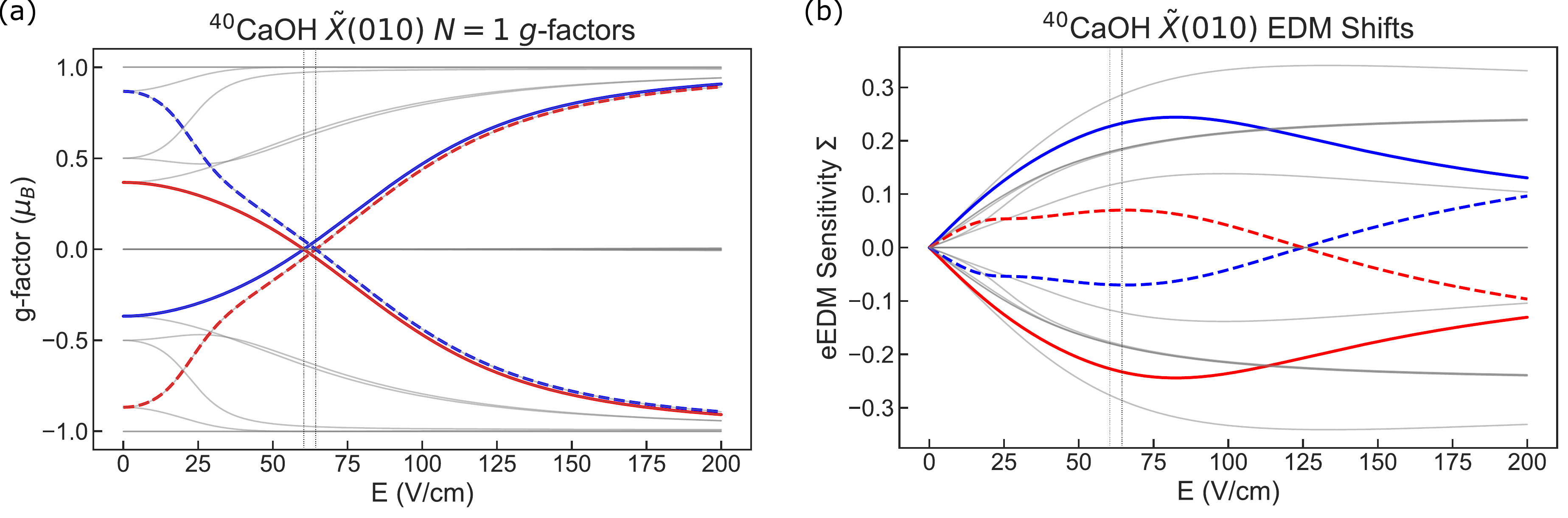}}
    \caption{Electric field tuning of $N=1$ zero \textit{g}-factor states near $B_Z=0$ in the absence of trap shifts. Blue lines denote $M_F=+1$ states and red lines $M_F=-1$. Solid traces denote the $J=1/2$ state pair and dashed traces denote the $J=3/2$ pair. The dotted vertical lines mark the electric field value of the zero \textit{g}-factor crossing without trap shifts, $\approx$60.5 V/cm for $J=1/2$ and $\approx$64.4  V/cm for $J=3/2$. Grayed out traces are other states in the $N=1$ manifold. (a) The \textit{g}-factor $g_S \mu_B \langle M_S \rangle$ as a function of the applied electric field. (b) eEDM sensitivity $\langle \Sigma \rangle$ as a function of the applied electric field. A consequence of the Hund's case (b) coupling scheme is that $\Sigma$ asymptotes to a maximum magnitude of $S/(N(N+1)) = 1/4$ for fields where the parity doublets are fully mixed but rotational mixing is negligible~\cite{Petrov2022YbOHEField}. For fields where $J$ is not fully mixed, some states can exhibit $|\Sigma| > 1/4$.}
     \label{figS1} 
\end{figure}

\begin{figure}[t]
    \resizebox{\textwidth}{!}{\includegraphics{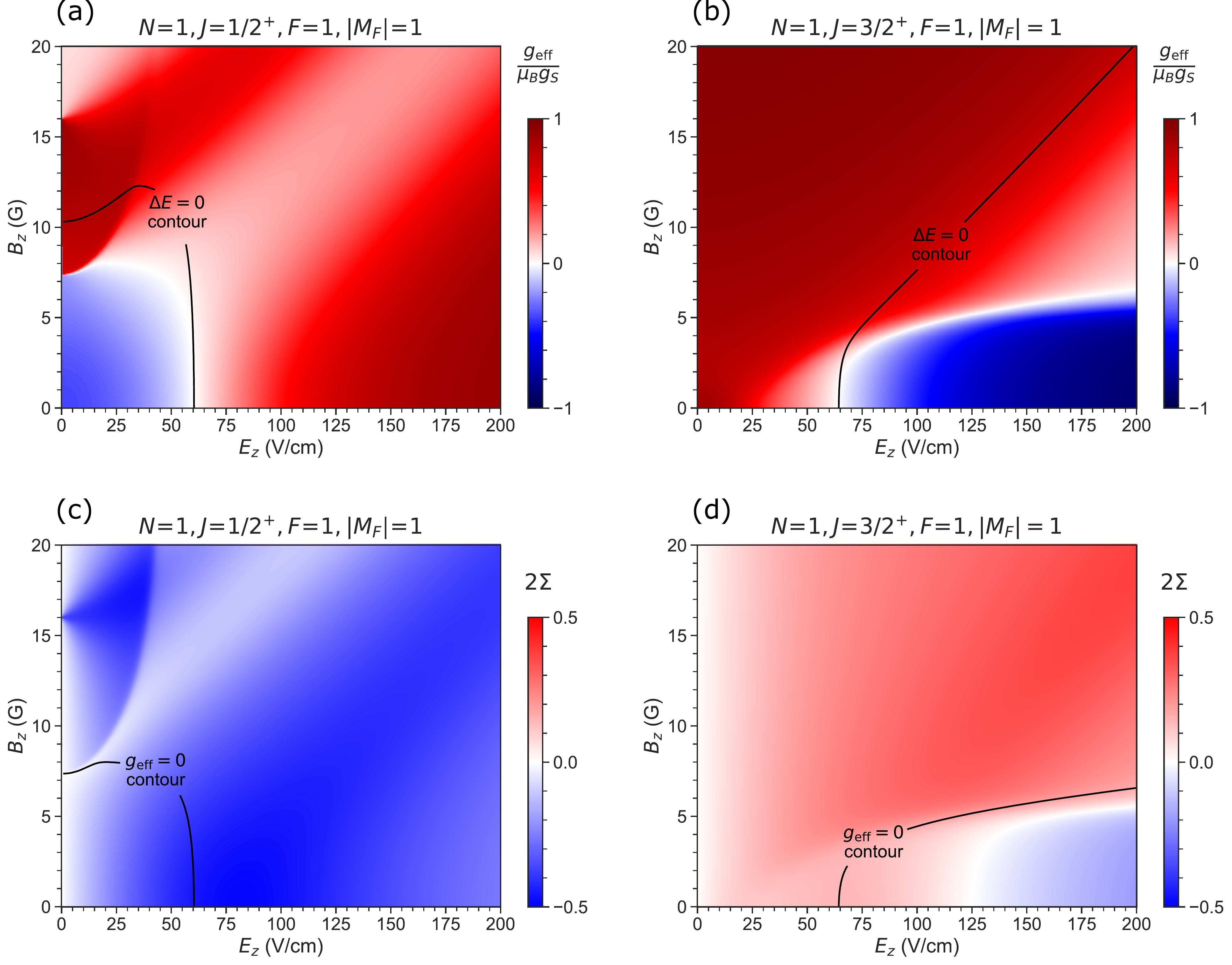}}
    \caption{Full electric and magnetic characterization of zero \textit{g}-factor states in the $N=1$ manifold of CaOH, without trap shifts. (a, b) 2D plots of the effective \textit{g}-factor difference between two $M=\pm1$ states, defined by $g_\text{eff} = g_S \mu_B \left(\langle M_S \rangle_{M=+1} - \langle M_S \rangle_{M=-1}\right)$. The plotted \textit{g}-factor is normalized by $g_S \mu_B$. The black line represents the contour where the $M=\pm1$ levels are nominally degenerate. (c, d) 2D plots of the eEDM sensitivity, $\langle \Sigma \rangle_{M=+1} - \langle \Sigma \rangle_{M=-1}$. The black line represents the $g_\text{eff} = 0$ contour.}
     \label{figS2} 
\end{figure}

\subsection{Characterization}

To locate zero \textit{g}-factor crossings and calculate eEDM sensitivities,  we model the $\widetilde{X}(010)$ level structure using an effective Hamiltonian approach~\cite{brown2003rotational,Hirota1985,merer1971rotational}:

\begin{subequations}
    \begin{align}
    H_\text{eff} & = H_\text{Rot} + H_\text{SR} + H_{\ell} + H_\text{Hyp} + H_\text{Zeeman} + H_\text{Stark} + H_\text{ODT}
    \label{eq:Heff}\\
        H_\text{Rot} & = B \left(\vec{N}^2-\ell^2\right)\\
        H_\text{SR} & = \gamma \left(\vec{N}\cdot \vec{S} - N_z S_z\right)\\
        H_\ell & = -q_{\ell} \left(N_+^2 e^{-i 2 \phi} + N_-^2 e^{i2 \phi}\right)\\
        H_\text{Hyp} & = b_F \vec{I}\cdot \vec{S} + \frac{c}{3} \left(3I_z S_z - \vec{I} \cdot \vec{S}\right)\\
        H_\text{Zeeman} & = g_S \mu_B B_Z S_Z\\
        H_\text{Stark} & = - \mu_{Z} E_Z\\
        H_\text{ODT} & = -\vec{d}\cdot \vec{E}_\text{ODT}
    \end{align}
\end{subequations}
Here, we use a similar Hamilton as Ref.~\cite{kozyryev2021enhanced}. $H_\text{Rot}$ is the rotational energy; $H_\text{SR}$ is the spin-rotation interaction accurate for low-$N$ bending mode levels, with $z$ defined in the molecule frame; $H_\ell$ is the $\ell$-type doubling Hamiltonian, with $\pm$ defined in the molecule frame, $\phi$ as the nuclear bending coordinate, and using the same phase convention as Ref.~\cite{Brown2003}; $H_\text{Hyp}$ is the hyperfine Fermi-contact and dipolar spin interactions, defined in the molecule frame; $H_\text{Zeeman}$ describes the interaction of the electron spin magnetic moment with the lab-frame magnetic field; $H_\text{Stark}$ is the interaction of the $Z$-component of molecule-frame electric dipole moment $\mu_E$ with the lab frame DC electric field, $E_Z$; and $H_\text{ODT}$ is the interaction of the molecular dipole moment operator $\vec{d}$ with the electric field of the ODT laser, $\vec{E}_\text{ODT} = \mathcal{E}_0/2(\hat{\epsilon}_\text{ODT}e^{-i\omega t}+\text{c.c.})$. 

To evaluate the molecule frame matrix elements, we follow the techniques outlined in Refs.~\cite{brown2003rotational,Hirota1985} to transform into the lab frame. The field-free Hamiltonian parameters are taken from Ref.~\cite{li1995bending}, except for the hyperfine parameters, which were determined by the observed line positions to be $b_F=2.45$ MHz and $c=2.6$ MHz, similar to those of the $\widetilde{X}(000)$ state \cite{scurlock1993hyperfine}. We use the same dipole moment, $|\mu|=1.47$~D, as the $\widetilde{X}(000)$ state, determined in Ref.~\cite{steimle1992supersonic}. Matrix elements of $H_\text{ODT}$ are calculated following Ref.~\cite{caldwell2020sideband} using the 1064nm dynamic polarizabilities reported in Ref.~\cite{hallas2022optical}.

For the calculations discussed below and in the main text, the ODT is polarized along the laboratory $Z$ axis and the molecules sit at a fixed trap depth of $160$~$\mu$K (corresponding to the average trap intensity seen by the molecules in the experiment).
As detailed in the main text, when the trapping light is aligned with $E_Z$, it acts like a weak electric field, shifting the zero \textit{g}-factor crossing by $\sim$ 1 V/cm from the field-free value. If the trapping light polarization is rotated relative to $E_Z$, tensor light shifts can couple states with $\Delta M_F = \pm 2$ or $\pm1$ (the linearity of the light ensures there are no $\Delta M_F=\pm 1$ vector shifts)~\cite{caldwell2020sideband}. The effects of this coupling are similar to those of transverse magnetic fields, which we discuss below. 

In the current work, we ignore nuclear and rotational Zeeman effects. Specifically, the magnetic sensitivity of CaOH receives small contributions from nuclear spin of the H atom and the rotational magnetic moment of both the electrons and the nuclear framework. While they have not yet been fully characterized, all of these effects will contribute at the $10^{-3} \mu_B$ level or less. These additional \textit{g}-factors do not depend strongly on the applied electric field, and result in a small shift of the zero \textit{g}-factor crossing location. Future work characterizing rotational magnetic moments of $\widetilde{X}(010)$ states of laser-coolable metal hydroxides can enable more accurate predictions of zero \textit{g}-factor field values.

In CaOH, each rotational state $N$ supports multiple $M=\pm1$ pairs of zero \textit{g}-factor states. The states at finite electric field can be labeled in terms of their adiabatically correlated zero-field quantum numbers $|N,J^p,F,M\rangle$. In the presence of trap shifts, the zero \textit{g}-factor states for $N=1$ occur at $E= 59.6$ V/cm for $|J=1/2^+,F=1,M=\pm1\rangle$ and at $E= 64.1$ V/cm for $|J=3/2^+,F=1,M=\pm1\rangle$. The $J=1/2, M=1$ state is a superposition of $47\%$ $M_N\ell=-1$, $50\%$ $M_N\ell=0$, and $3\%$ $M_N\ell=1$, while the $J=3/2, M=1$ state is $43\%$ $M_N\ell=-1$, $48\%$ $M_N=0$, and $9\%$ $M_N \ell=1$. Both states are weak-electric-field seekers, yet the opposite molecule frame orientation of the spin results in differences in the value of $\Sigma$ and the \textit{g}-factor slope.  For CaOH, the magnetic sensitivity and eEDM sensitivity of $N=1$ zero \textit{g}-factor states are shown in Fig.~\ref{figS1}.

By diagonalizing $H_\text{eff}$ over a grid of $(E_Z, B_Z)$ values, we can obtain 2D plots of \textit{g}-factors and eEDM sensitivities shown in Fig.~\ref{figS2}. For generality, we consider the molecular structure in the absence of trap shifts. Using the $Z$-symmetry of the Hamiltonian, we separately diagonalize each $M_F$ block to avoid degeneracies at $B_Z=0$. Continuous 2D surfaces for eigenvalues and eigenvectors are obtained by ordering eigenstates at each value of $(E,B)$ according to their adiabatically correlated free field state. The application of an external magnetic field parallel to the electric field results in $\langle M_S \rangle \neq 0$ for an individual zero \textit{g}-factor state, but the differential value between a zero \textit{g}-factor pair can still have $\Delta \langle M_S\rangle=0$. This differential value means the superposition of a zero \textit{g}-factor pair can maintain magnetic insensitivity and EDM sensitivity over a range of fields, for example up to $\sim$5 G for the $J=1/2, N=1$ pair. 

The procedure we use here for identifying zero \textit{g}-factor states can be generically extended to searching for favorable transitions between states with differing eEDM sensitivities, similar to what has been already demonstrated in a recent proposal to search for ultra-light dark matter using SrOH~\cite{kozyryev2021enhanced}. In addition, there are also fields of $B_Z\approx 10 - 20$ G and $E_Z\approx0$ where opposite parity states are tuned to near degeneracy. This is the field regime that has been proposed for precision measurements of parity-violation in optically trapped polyatomic molecules~\cite{norrgard2019nuclear}.

We note that zero \textit{g}-factor pairs also occur in $N=2^-$. The crossings occur around $400-500$ V/cm for states correlated with the negative parity manifold. Since many interactions increase in magnitude with larger $N$, the overall electric field scale of the intermediate regime increases. Additionally, the robustness of zero \textit{g}-factor states also improves, with some pairs able to  maintain $\Delta \langle M_S\rangle=0$ for magnetic fields up to $40$ G. These $N=2$ pairs also have non-zero eEDM sensitivity for a wide range of magnetic field values. 

\section{Transverse magnetic fields}

\subsection{Transverse Field Sensitivity}

\begin{figure}[t]
    \resizebox{0.5\textwidth}{!}{\includegraphics{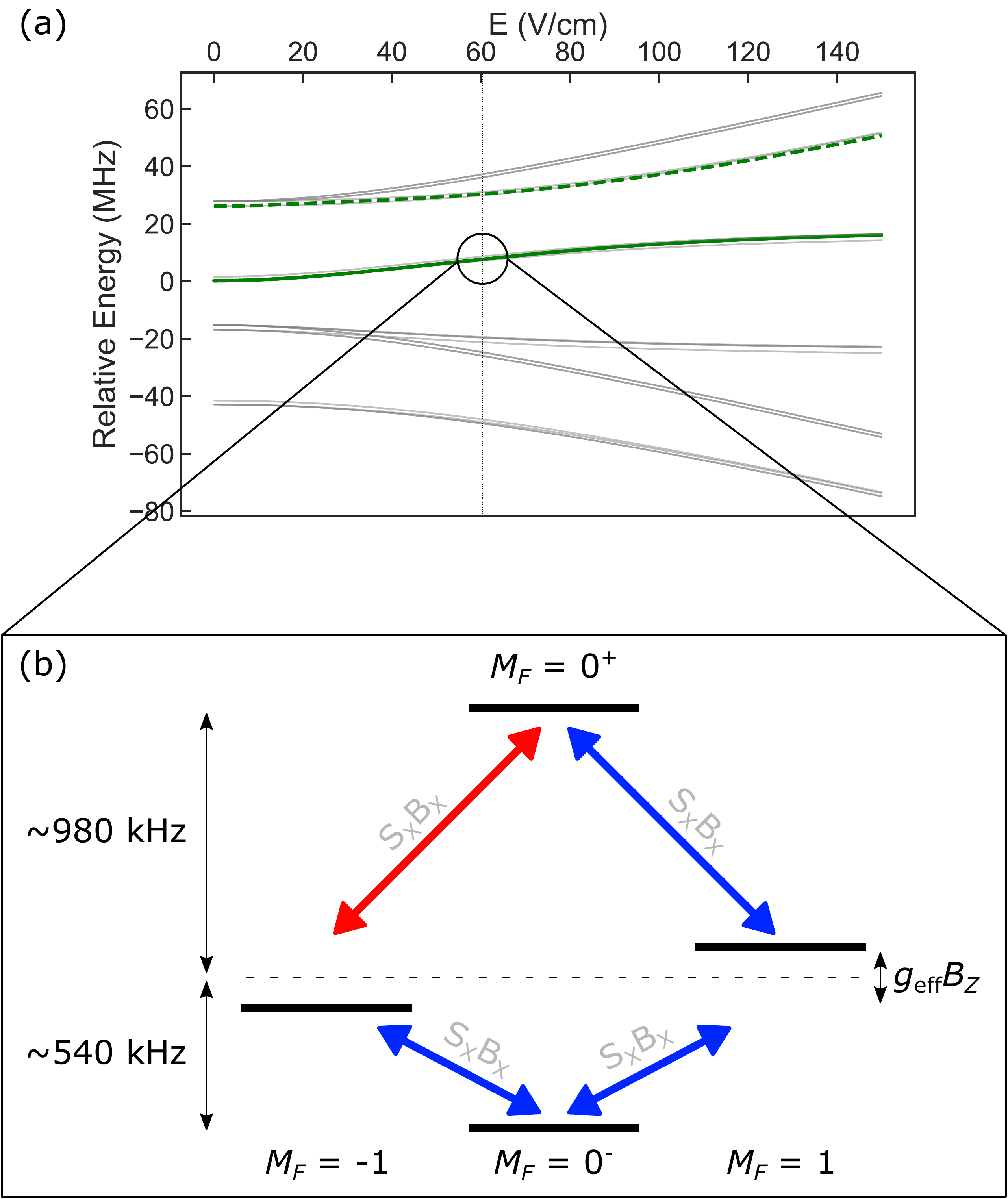}}
    \caption{(a) Stark shifts for $N=1$ in CaOH. The $J=1/2^+$ zero \textit{g}-factor states are shown with a solid green line, while the $J=3/2^+$ zero \textit{g}-factor states are indicated with a dashed green line. All other levels are grayed out. A vertical dotted line indicates the location of the $J=1/2^+$ zero \textit{g}-factor crossing. (b) A zoomed in level diagram of the $J=1/2^+$ zero \textit{g}-factor hyperfine manifold. The bias field splitting $g_\text{eff} B_Z$ is not to scale. Transverse field couplings are shown with double sided arrows, with blue (red) indicating negative (positive) $S_X$ matrix element.}
     \label{figS3} 
\end{figure}

We now expand our discussion to include the effect of transverse magnetic fields. Their effects can by modeled by adding $B_X S_X$ and $B_Y S_Y$ terms to the effective Hamiltonian, which have the selection rule $\Delta M_F = \pm 1$. For this discussion, we focus on the level structure of the $N=1, J=1/2^+$ manifold in CaOH near the zero \textit{g}-factor crossing at 60.5 V/cm in the absence of trap shifts, shown in Figure~\ref{figS3}. We note if there were no nuclear spin $I$, the two zero \textit{g}-factor states would be $M_J=\pm 1/2$ states separated by $\Delta M=1$. In such a case these degenerate states would be directly sensitive to transverse fields at first order, thereby reducing the \textit{g}-factor suppression. 

Due to the hyperfine structure from the nuclear spin of the H atom in CaOH, the degenerate $M_F=\pm1$ states in a zero \textit{g}-factor pair are coupled by second order transverse field interactions. These interactions are mediated via the $M_F=0^\pm$ states, where $\pm$ denotes the upper or lower states. Using a Schrieffer–Wolff (aka Van-Vleck) transformation, we can express the effective Hamiltonian matrix for second order coupling between the $M_F=\pm1$ states. We write the states as $| M_F \rangle$, and for convenience we take the transverse field to point along $X$:

\begin{equation}\label{eq:Htrans}
    H_{+1,-1}  = -(g_S \mu_B B_X)^2 \left(\frac{\langle -1|S_X|0^+\rangle\langle0^+| S_X |+1\rangle}{\Delta E_{0^+}} 
    + \frac{\langle-1|S_X|0^-\rangle\langle 0^-| S_X |+1\rangle}{\Delta E_{0^-}} \right)
\end{equation}\\
Here, $\Delta E_{0^\pm}$ is the energy difference of the $M_F=0^\pm$ levels from the $M_F=\pm 1$ levels. Our model provides the following values: $\langle 0^-|S_X|+1\rangle=\langle 0^-|S_X|-1\rangle=-0.18$, $\langle 0^+|S_X|+1\rangle=-0.16$, and $\langle 0^+|S_X|-1\rangle=0.16$. The difference in sign is a result of Clebsh-Gordon coefficient phases, and only the relative phase is relevant. We also have $\Delta E_{0^+} = 0.98$ MHz and $\Delta E_{0^-}= -0.54$ MHz. The combination of phases precludes the possibility of destructive interference. With these parameters and defining $g_\perp = H_{+1,-1}/B_X$, then eqn.~\ref{eq:Htrans} evaluates to $(g_S\mu_B B_X)^2 (0.086/$MHz$) \approx (0.68$ MHz/G$^2$)$ B_X^2$. Our model estimates the transverse sensitivity at $B_X \sim 1$ mG to be $g_\perp \mu_B \sim 7\times10^{-4}$ MHz/G, of the same order as the neglected nuclear and rotational Zeeman terms. The suppressed transverse field sensitivity bounds the magnitude of $B_Z$, which must be large enough to define a quantization axis for the spin, $g_\text{eff} B_Z \gg g_\perp B_\perp$. 

\subsection{Cancellation of transverse magnetic fields}

When transverse magnetic fields are dominant, the electron will be quantized along the transverse axis and there is minimal spin precession by the bias $B_Z$ field. The transverse coupling results in eigenstates given by $(|M_F=1\rangle \pm e^{i\phi}|M_F=-1\rangle)/\sqrt{2}$, where the phase $\phi$ is set by the direction of $\vec{B}$ in the transverse plane. If $\phi = 0$ or $\pi$, only one of these states is bright to the $\hat{X}$-polarized state preparation microwaves, which means the initial state is stationary under the transverse fields. For all other orientations, the transverse field causes spin precession with varying contrast, depending on the specific value of $\phi$. 

We are able to use transverse spin precesion to measure and zero transverse fields to the mG level. We do so by operating with minimal bias field $B_Z\approx 0$ and operating $E_Z$ near the zero \textit{g}-factor crossing, such that $g_\text{eff} B_Z < g_\perp B_\perp$. We then apply a small transverse magnetic field to perform transverse spin precession. Here, the dynamics are dominated by the transverse fields rather than the $Z$ fields. We obtain field zeros by iteratively minimizing the precession frequency by tuning the bias fields $B_X$ and $B_Y$.

\section{Imperfect Field Reversal}

We briefly present a systematic effect involving non-reversing fields in eEDM measurements with zero \textit{g}-factor states and discuss methods for its mitigation. The electric field dependence of $g_\text{eff}$ can mimic an eEDM signal when combined with other systematic effects, very much like in $^3\Delta_1$ molecules~\cite{acme2018improved,cairncross2017precision}. When the sign of $E_Z$ is switched, a non-reversing electric field $E_\text{NR}$ will cause a \textit{g}-factor difference of $g_\text{NR}=(\mathrm{d}{g_\text{eff}}/\mathrm{d}{E_Z})  E_\text{NR}$. This will give an additional spin precession signal $g_\text{NR} B_Z$. By perfectly reversing $B_Z$ as well, this precession signal can be distinguished from a true EDM signal. However, if there is also a non-reversing magnetic field $B_\text{NR}$, there will still be a residual EDM signal given by $(\mathrm{d}{g}/\mathrm{d}{E})  E_\text{NR} B_\text{NR}$. Using the measured slope of $\sim$0.03 (MHz/G)/(V/cm), and using conservative estimates of $E_\text{NR}\sim 1$ mV/cm and $B_\text{NR}\sim 1$ $\mu$G, we obtain an estimate precession frequency of $\sim$30 $\mu$Hz. While this is an order of magnitude smaller than the statistical error for the current best eEDM measurement measurement~\cite{lasner2019order}, it is still desirable to devise methods to reduce the effect further.

Performing eEDM measurements at different zero \textit{g}-factor states can help suppress systematic errors resulting from the above mechanism. For example, the $N=1, J=3/2$ zero crossing has a different magnitude for $\Sigma$, which can be used to distinguish a true eEDM from a systematic effect. Both $N=1$ crossings are only separated by $\sim$4~V/cm. Furthermore, the zero \textit{g}-factor states in $N=2^-$ can also be used for systematic checks, as they additionally offer different $g_\text{eff}$ vs $E_Z$ slopes as well as different $\Sigma$ values. The $N=2^-$ states can be populated directly by the photon-cycling used to pump into the bending mode. 

\section{Spin Precession Near Zero \textit{g}-Factor}

\begin{figure}
\centering
    \resizebox{0.6\textwidth}{!}{\includegraphics{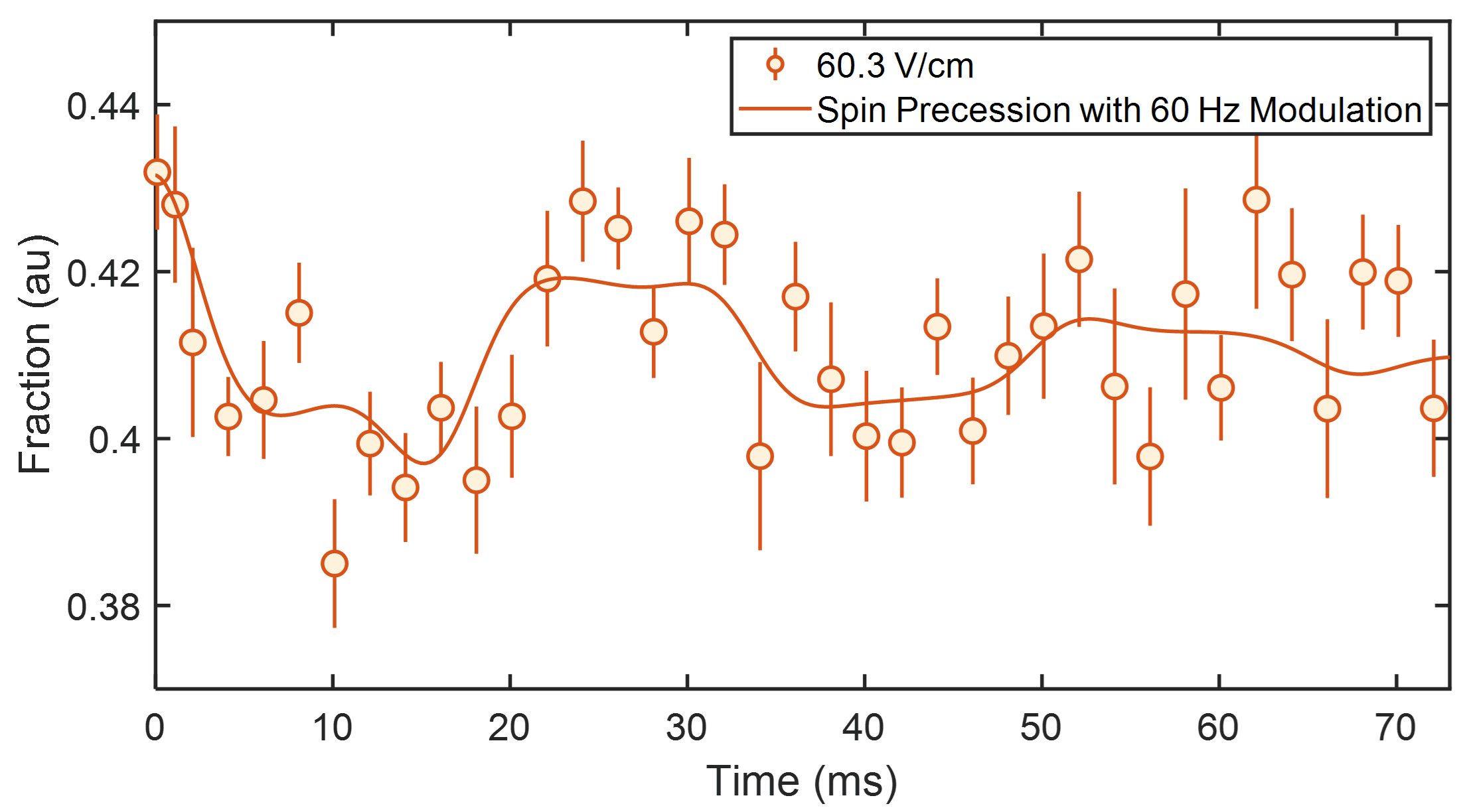}}
    \caption{Spin precession at $E_Z = 60.3$~V/cm and $B_Z = 2$~mG. The fit includes a 60 Hz time-varying magnetic field whose amplitude and phase are measured with a magnetometer. The coherence time fits to 30 ms.}
     \label{figS4} 
\end{figure}

As discussed in the main text, the longest achievable coherence times occur at at combination of low effective $g$-factors (which suppress $\delta B_Z$ decoherence) and low magnetic bias fields (which suppress $\delta \mu_\text{eff}$ decoherence). These low $g$-factors and bias fields only very weakly enforce a quantization axis along $Z$, enhancing the potential for transverse magnetic fields $B_\perp$ to contribute. Such fields have the effect of (a) reducing the spin precession contrast and (b) altering the observed precession frequency. To avoid these effects, the condition $g_\text{eff} B_Z > g_\perp B_\perp$ must therefore be satisfied. To achieve this, we zero the transverse magnetic fields by intentionally taking spin precession data at $B_Z \approx 0$ and $g_\text{eff}\approx 0$ while varying the transverse fields $B_X$ and $B_Y$. By minimizing the spin precession frequency as a function of the transverse fields, we reduce $B_\perp$ to approximately 1 mG. In addition, long-term drifts in the dc magnetic field along all three axes are compensated by actively feeding back on the magnetic field as measured with a fluxgate magnetometer. Under these conditions, at an electric field of 60.3 V/cm (corresponding to $\mu_\text{eff}=0.02$ MHz/G) and a bias field of $B_Z\approx2$ mG, we achieve a coherence time of 30 ms (Fig. \ref{figS4}).

At these very low bias fields, the molecules are also sensitive to 60 Hz magnetic field noise present in the unshielded apparatus, whose amplitude is on the same order as $B_Z$. Since the experiment is phase stable with respect to the AC line frequency, this 60 Hz magnetic field fluctuation causes a time-dependent spin precession frequency. A fluxgate magnetometer is used to measure the amplitude and phase of this 60 Hz field, which are then used as fixed parameters in the fit shown in Figure S4.


\clearfmfn

\end{document}